\documentclass[prl,twocolumn,superscriptaddress,showpacs]{revtex4}\newif\ifpdf

\usepackage{dcolumn}% Align table columns on decimal point
\usepackage{bm}% bold math
\usepackage{amsmath} % Typical maths resource packages

\ifx\pdfoutput\undefined
  \pdffalse % we are not running PDFLaTeX
\else
  \pdfoutput=1 % we are running PDFLaTeX
  \pdftrue
\fi

\ifpdf
  \usepackage[pdftex]{graphicx}
  \usepackage[pdftex,dvipsnames,usenames]{color}\else
  \usepackage{graphicx}
  \usepackage[dvips,dvipsnames,usenames]{color}\fi

\begin{document}\ifpdf
  \DeclareGraphicsExtensions{.pdf, .jpg, .tif}
\else
  \DeclareGraphicsExtensions{.eps}
\fi

% You should use BibTeX and apsrev.bst for references
% Choosing a journal automatically selects the correct APS
% BibTeX style file (bst file), so only uncomment the line
% below if necessary.
%\bibliographystyle{apsrev}

%\begin{document}


\title{\boldmath Measurement of ${\cal B}(t\rightarrow Wb)/
{\cal B}(t\rightarrow Wq)$ at the Collider Detector at Fermilab}



% Official author list modified for REVTeX 4!
\affiliation{Institute of Physics, Academia Sinica, Taipei, Taiwan 11529, Republic of China}  
\affiliation{Argonne National Laboratory, Argonne, Illinois 60439}  
\affiliation{Institut de Fisica d'Altes Energies, Universitat Autonoma de Barcelona, E-08193, Bellaterra (Barcelona), Spain}  
\affiliation{Istituto Nazionale di Fisica Nucleare, University of Bologna, I-40127 Bologna, Italy}  
\affiliation{Brandeis University, Waltham, Massachusetts 02254}  
\affiliation{University of California, Davis, Davis, California  95616}  
\affiliation{University of California, Los Angeles, Los Angeles, California  90024}  
\affiliation{University of California, San Diego, La Jolla, California  92093}  
\affiliation{University of California, Santa Barbara, Santa Barbara, California 93106}  
\affiliation{Instituto de Fisica de Cantabria, CSIC-University of Cantabria, 39005 Santander, Spain}  
\affiliation{Carnegie Mellon University, Pittsburgh, PA  15213}  
\affiliation{Enrico Fermi Institute, University of Chicago, Chicago, Illinois 60637}  
\affiliation{Joint Institute for Nuclear Research, RU-141980 Dubna, Russia}  
\affiliation{Duke University, Durham, North Carolina  27708}  
\affiliation{Fermi National Accelerator Laboratory, Batavia, Illinois 60510}  
\affiliation{University of Florida, Gainesville, Florida  32611}  
\affiliation{Laboratori Nazionali di Frascati, Istituto Nazionale di Fisica Nucleare, I-00044 Frascati, Italy}  
\affiliation{University of Geneva, CH-1211 Geneva 4, Switzerland}  
\affiliation{Glasgow University, Glasgow G12 8QQ, United Kingdom} 
\affiliation{Harvard University, Cambridge, Massachusetts 02138}  
\affiliation{Division of High Energy Physics, Department of Physics, University of Helsinki and Helsinki Institute of Physics, FIN-00014, Helsinki, Finland} 
\affiliation{Hiroshima University, Higashi-Hiroshima 724, Japan}  
\affiliation{University of Illinois, Urbana, Illinois 61801}  
\affiliation{The Johns Hopkins University, Baltimore, Maryland 21218}  
\affiliation{Institut f\"{u}r Experimentelle Kernphysik, Universit\"{a}t Karlsruhe, 76128 Karlsruhe, Germany}  
\affiliation{High Energy Accelerator Research Organization (KEK), Tsukuba, Ibaraki 305, Japan}  
\affiliation{Center for High Energy Physics: Kyungpook National University, Taegu 702-701; Seoul National University, Seoul 151-742; and SungKyunKwan University, Suwon 440-746; Korea}  
\affiliation{Ernest Orlando Lawrence Berkeley National Laboratory, Berkeley, California 94720}  
\affiliation{University of Liverpool, Liverpool L69 7ZE, United Kingdom}  
\affiliation{University College London, London WC1E 6BT, United Kingdom}  
\affiliation{Massachusetts Institute of Technology, Cambridge, Massachusetts  02139}  
\affiliation{Institute of Particle Physics: McGill University, Montr\'{e}al, Canada H3A~2T8; and University of Toronto, Toronto, Canada M5S~1A7}  
\affiliation{University of Michigan, Ann Arbor, Michigan 48109}  
\affiliation{Michigan State University, East Lansing, Michigan  48824}  
\affiliation{Institution for Theoretical and Experimental Physics, ITEP, Moscow 117259, Russia}  
\affiliation{University of New Mexico, Albuquerque, New Mexico 87131}  
\affiliation{Northwestern University, Evanston, Illinois  60208}  
\affiliation{The Ohio State University, Columbus, Ohio  43210}  
\affiliation{Okayama University, Okayama 700-8530, Japan} 
\affiliation{Osaka City University, Osaka 588, Japan}  
\affiliation{University of Oxford, Oxford OX1 3RH, United Kingdom}  
\affiliation{University of Padova, Istituto Nazionale di Fisica Nucleare, Sezione di Padova-Trento, I-35131 Padova, Italy}  
\affiliation{University of Pennsylvania, Philadelphia, Pennsylvania 19104}  
\affiliation{Istituto Nazionale di Fisica Nucleare Pisa, Universities of Pisa, Siena and Scuola Normale Superiore, I-56127 Pisa, Italy}  
\affiliation{University of Pittsburgh, Pittsburgh, Pennsylvania 15260}  
\affiliation{Purdue University, West Lafayette, Indiana 47907}  
\affiliation{University of Rochester, Rochester, New York 14627}  
\affiliation{The Rockefeller University, New York, New York 10021}  
\affiliation{Istituto Nazionale di Fisica Nucleare, Sezione di Roma 1, University di Roma ``La Sapienza," I-00185 Roma, Italy} 
\affiliation{Rutgers University, Piscataway, New Jersey 08855}  
\affiliation{Texas A\&M University, College Station, Texas 77843}  
\affiliation{Texas Tech University, Lubbock, Texas 79409}  
\affiliation{Istituto Nazionale di Fisica Nucleare, University of Trieste/\ Udine, Italy}  
\affiliation{University of Tsukuba, Tsukuba, Ibaraki 305, Japan}  
\affiliation{Tufts University, Medford, Massachusetts 02155}  
\affiliation{Waseda University, Tokyo 169, Japan}  
\affiliation{Wayne State University, Detroit, Michigan  48201}  
\affiliation{University of Wisconsin, Madison, Wisconsin 53706}  
\affiliation{Yale University, New Haven, Connecticut 06520}  
\author{D.~Acosta}
\affiliation{University of Florida, Gainesville, Florida  32611} 
\author{J.~Adelman}
\affiliation{Enrico Fermi Institute, University of Chicago, Chicago, Illinois 60637} 
\author{T.~Affolder}
\affiliation{University of California, Santa Barbara, Santa Barbara, California 93106} 
\author{T.~Akimoto}
\affiliation{University of Tsukuba, Tsukuba, Ibaraki 305, Japan} 
\author{M.G.~Albrow}
\affiliation{Fermi National Accelerator Laboratory, Batavia, Illinois 60510} 
\author{D.~Ambrose}
\affiliation{Fermi National Accelerator Laboratory, Batavia, Illinois 60510} 
\author{S.~Amerio}
\affiliation{University of Padova, Istituto Nazionale di Fisica Nucleare, Sezione di Padova-Trento, I-35131 Padova, Italy} 
\author{D.~Amidei}
\affiliation{University of Michigan, Ann Arbor, Michigan 48109} 
\author{A.~Anastassov}
\affiliation{Rutgers University, Piscataway, New Jersey 08855} 
\author{K.~Anikeev}
\affiliation{Fermi National Accelerator Laboratory, Batavia, Illinois 60510} 
% ...

\collaboration{CDF Collaboration}
\noaffiliation

\date{\today}

\newcommand{\qqbar}{\ensuremath{q\bar q}}
\newcommand{\Wenu}{\ensuremath{W\rightarrow e \nu}}
\newcommand{\Zee}{\ensuremath{Z\rightarrow e^{+}e}^{-}}
\newcommand{\met}{\ensuremath{E\kern-0.6em\lower-.1ex\hbox{/}_T}}
\newcommand{\metx}{\ensuremath{E\kern-0.6em\lower-.1ex\hbox{/}_{T_x}}}

\begin{abstract}
  We present a measurement of the ratio of top-quark branching
  fractions $R= {\cal B}(t\rightarrow Wb)/{\cal B}(t\rightarrow Wq)$,
  where $q$ can be a $b$, $s$ or a $d$ quark,
  using lepton-plus-jets and dilepton data sets with integrated
  luminosity of $\sim$162~pb$^{-1}$ collected with the Collider Detector at
  Fermilab during Run~II of the Tevatron.  The measurement is derived
  from the relative numbers of $t\bar{t}$ events with different multiplicity of
  identified secondary vertices.
  %in  which the $t\bar{t}$ background has been determined as a function
  %the secondary vertex multiplicity.
  We set a lower limit of $R>0.61$ at 95\% confidence level.

\vspace{0.5cm}  
\end{abstract}

\pacs{14.65.Ha, 12.15.Hh}
%\keywords{}

\maketitle

The top quark as described by the Standard Model (SM) is expected to
decay to a $W$ boson and a bottom quark at least 99.8\% of the time at
90\% confidence level (CL)~\cite{pdg}.  The Cabbibo-Kobayashi-Maskawa
(CKM) quark-mixing matrix~\cite{c,km} element $|V_{tb}|$ is expected
to be very close to unity from the assumption of a unitary,
three-generation matrix and the measured small values of $|V_{ub}|$
and $|V_{cb}|$~\cite{pdg}.  A measurement of the ratio of top-quark
branching fractions $R = {\cal B}(t\rightarrow Wb)/{\cal
  B}(t\rightarrow Wq)$, where $q$ can be a $b$, $s$ or a $d$ quark,
significantly less than unity would contradict our current theoretical
assumptions, implying either non-SM top decay, a non-SM background to
top-pair production, or a fourth generation of quarks.  A previous
measurement has set a lower limit of $R > 0.56$ at 95\%
CL~\cite{oldR}.  In this Letter we present a measurement of $R$ using
$t\bar{t}$ events collected at the Collider Detector at Fermilab (CDF)
during Run~II of the Tevatron, a proton-antiproton collider with
center of mass energy of $\sqrt{s}=1.96$~TeV.  The integrated
luminosity of the data sample used in this analysis is
$\sim$162~pb$^{-1}$.

Our measurement uses $t\bar{t}$-pair events.  The
lifetime of top is too short for hadronization to occur, and the SM
strongly favors an essentially immediate decay of each quark to a real
$W$ boson and weak-isospin -1/2 quark; if $R=1$, this is always a $b$
quark.  To maintain high detection and trigger efficiencies and low
background levels, we only consider $t\bar{t}$ final states in which
at least one $W$ has decayed leptonically.  Events in which one $W$
decays leptonically are called ``lepton-plus-jets'' (L+J) events, and
events with two leptonic decays are called ``dilepton'' (DIL) events.
Values of $R$ are determined separately for each of these sets of
events, and are combined in the end to set a lower limit on $R$.  The
greater statistical power comes from the L+J sample.

The measurement requires both the counting of $b$-quark jets and the
determination of the $t\bar{t}$ content as a function of the $b$-quark
multiplicity.  We identify (``tag'') $b$-quark jets by identifying
displaced secondary vertices using the SECVTX algorithm~\cite{l+j}.
$R$ is extracted from the relative rates of events with zero, one, and
two tags; any two rates determine $R$ uniquely, while all three rates
jointly overdetermine $R$.  A novel feature of this measurement is the
inclusion of the 0-tag L+J event rate, which is determined using event
kinematics and an artificial neural net (ANN) technique.  As $R$
depends only on relative rates, this measurement is independent of any
assumptions of the overall $t\bar{t}$ cross section.  However, our
measurement of $R$ does depend critically on the knowledge of the
efficiency to identify $b$ jets.  To extract $R$ we use the efficiency
to tag jets in $t\bar{t}$ events estimated with a Monte Carlo (MC)
sample in which tagging efficiencies have been tuned to match jet
data~\cite{l+j}.

The CDF detector for Run~II~\cite{cdfII_new} consists of a
charged-particle tracking system in a magnetic field of 1.4 T,
segmented electromagnetic and hadronic 
calorimeters and muon detectors.  A silicon microstrip detector
provides tracking over the radial range 1.5 to 28~cm, and is essential
for the detection of displaced secondary vertices.  
%A 3.1-m-long
%open-cell drift chamber covers the radial range from 40 to 137~cm.
The fiducial region of the silicon detector covers the pseudo-rapidity
range $|\eta| < 2$, while the central tracking system and muon 
chambers provide coverage for $|\eta| < 1$~\cite{geom}.  
%Segmented electromagnetic and hadronic
%sampling calorimeters surround the tracking system and measure the
%energy flow of interacting particles in the range $|\eta| < 3.6$.  A
%set of drift chambers located outside the central hadron calorimeters
%and another set behind a 60~cm-thick iron shield, accompanied with scintillation counters, detect energy
%deposition from muon candidates with $|\eta| \leq 0.6$.  Additional
%drift chambers and scintillation counters detect muons in the region
%$0.6 \leq |\eta| < 1.0$.  
A three-level trigger system is used to
select events with electron (muon) candidates with $E_T$ ($p_T$) $>
18$~GeV (18 GeV/$c$), which form the data set for this analysis.
%Trigger efficiencies for both L+J and DIL
%events are typically greater than 90\%.

L+J events consist of one isolated high-$p_T$ lepton ($e$ or
$\mu$), large missing transverse energy ($\met$) due to the undetected
neutrino, and four hadronic jets.  Two of these jets arise from the
hadronic decay of the other $W$, and the other two arise from the
top-daughter quarks $q$.  The L+J selection requirements are described
in detail elsewhere~\cite{l+j}.  Briefly, we require the presence of
an isolated lepton which has transverse momentum greater than
20~GeV/$c$, that $\met$ is at least 20~GeV, and that there is a
minimum of four jets, clustered with a cone-based algorithm having
cone dimension $\Delta R = \sqrt{(\Delta \phi)^2+(\Delta \eta)^2} =
0.4$, within $|\eta| < 2$ and with corrected transverse
energy~\cite{l+j} greater than 15~GeV.  These requirements select 107
events.

DIL events consist of two charged leptons ($ee$, $\mu\mu$ or
$e\mu$), large $\met$ due to the undetected neutrinos, and two jets
from the top-daughter quarks $q$.  The DIL selection requirements are
described in detail elsewhere~\cite{dil}.  Compared to the L+J
selection, we demand an additional lepton, but only a minimum of two
energy-corrected~\cite{dil} jets, with the same requirements as
before.  These requirements select 11 events.

Both event samples are subdivided on the basis of the number of
identified $b$ jets in the event.  The number of events in each
subsample with $i$ tagged jets are given in Table~\ref{yields}.  The
2-tag subsample is defined to include events with $\ge 2$ tags; in
this data sample we observe no events with more than two tagged jets.

\begin{table}
\centering
\caption{\label{yields} Summary of observed number of events with $i$~tags in
the L+J and DIL samples, with estimates of nominal $t\bar{t}$ event-tagging
efficiencies, background levels and expected event yields.  The
L+J 0-tag background is measured with an ANN.  The efficiency estimates
and the 1-tag and 2-tag L+J background estimates are given for $R=1$.  
Equations \ref{sig} and \ref{sig2} are used for the calculation of the 
expected total number of events $N_i^\text{exp}$. The statistical and 
systematic uncertainties have been combined.}
\begin{tabular*}{\columnwidth}{c@{\extracolsep{\fill}}c@{\extracolsep{\fill}}c@{\extracolsep{\fill}}c} \hline \hline
{\bf Lepton\,+\,Jets\,(L+J)} &{\bf 0-tag}& {\bf 1-tag} & {\bf 2-tag} \\ \hline
Efficiency 
($\epsilon_i(R=1)$) & 0.45 $\pm$ 0.03 & 0.43 $\pm$ 0.02 & 0.12 $\pm$ 0.02 \\
Background ($N_i^\text{bkg}$) & $62.4 \pm 9.0$ & $4.2\pm0.7$ & $0.2\pm0.1$ \\
Total expected ($N_i^\text{exp}$) &$80.4\pm 5.2$&$21.5\pm 4.1$ & $5.0\pm1.4$ \\
Observed ($N_i^\text{obs}$)  & 79     & 23  & 5 \\ \hline
%
%                      & \multicolumn{3}{c}{\bf Dileptons (DIL)} \\
{\bf Dileptons (DIL)}     & {\bf 0-tag} & {\bf 1-tag} & {\bf 2-tag} \\ \hline
Efficiency
($\epsilon_i(R=1)$)  & 0.47 $\pm$ 0.03 & 0.43 $\pm$ 0.02 & 0.10 $\pm$ 0.02 \\
Background ($N_i^\text{bkg}$) & $2.0\pm0.6$ & $0.2 \pm 0.1$  & $<0.01$ \\
Total expected ($N_i^\text{exp}$) &$6.1\pm0.4$&$4.0\pm 0.2$ & $0.9 \pm 0.2$ \\
Observed ($N_i^\text{obs}$)    & 5 & 4 & 2 \\ \hline \hline
\end{tabular*}
\end{table}

In the L+J sample, the dominant background is $W$ production in
association with jets from QCD processes (``$W$+jets'' events).  In
the 1-tag and 2-tag subsamples we make an {\it a priori} estimate of
the backgrounds with a collection of data-driven and simulation
techniques that are described in detail elsewhere~\cite{l+j}.  The
backgrounds in these subsamples include $W$ production in association
with heavy-flavor jets ($Wb\bar{b}$, $Wc\bar{c}$, $Wc$), $W$
production in association with light-flavor jets that are incorrectly
identified as $b$ jets (``mistags''), QCD multi-jet (``QCD'') events
containing fake or real leptons and/or incorrectly-measured $\met$,
dibosons ($WW$, $WZ$) and single-top quark production.  The background
estimate requires a small correction for $R \neq 1$.  The background
estimate for $R=1$ in these subsamples is given in Table~\ref{yields}.
The uncertainties on the estimate are dominated by uncertainties in
the fraction of $W$+jets events that include heavy flavor and on the
normalization of the QCD background rate.

By construction, the {\it a priori} method cannot predict the
background level in the 0-tag L+J sample, where the $W$+jets
production rate dominates that for $t\bar{t}$ pairs; instead we make
use of event kinematics~\cite{dima}.  The artificial neural
net~\cite{snns} is trained with the $t\bar{t}$ signal
(HERWIG~\cite{herwig}) and $W$+jets background
(HERWIG+ALPGEN~\cite{alpgen}) events simulated with a detailed
detector description based on GEANT~\cite{geant}.  There is an
additional QCD background which is modeled using data with
non-isolated leptons.  We find optimal signal to background
discrimination with an ANN structure of nine input variables, one
intermediate layer with ten nodes, and one output unit.  The variables
used are the transverse energies of the four leading jets, the minimum
di-jet mass, the di-jet transverse mass with value closest to the mass
of the $W$, the scalar sum of the transverse energies of all leptons
and jets, the total longitudinal momentum divided by the total
transverse momentum, and the event aplanarity.

\begin{figure}
\includegraphics[width=\columnwidth]{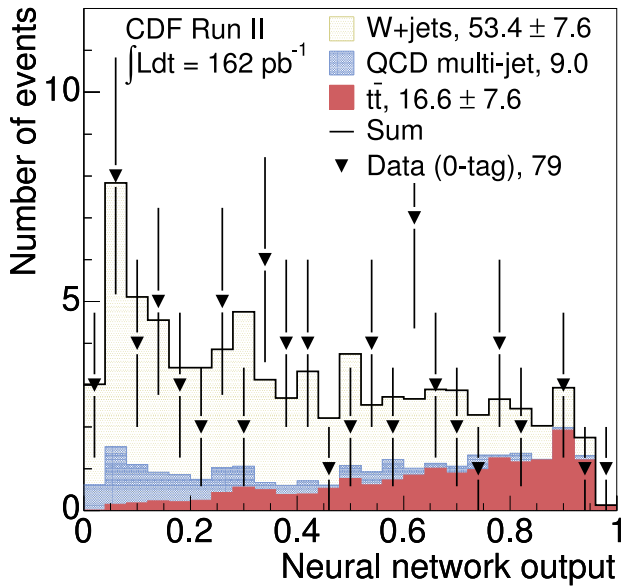}
\caption{Fit of the ANN output in the 0-tag L+J data set (triangles)
  with a sum of 3 components: $W$+jets (upper), QCD multi-jet
  (middle), and $t\bar t$ (lower). The QCD normalization is
  independently estimated and not varied in the fit; its
  shape is determined from the non-isolated lepton data.  }
\label{ann}
\end{figure}

The ANN output ranges from zero for background-like events to one for
signal-like events.  We perform a binned maximum likelihood fit of the
ANN output distribution for the $t\bar{t}$ fraction in the 0-tag
subsample.  The fraction of events from QCD backgrounds is fixed to
11.4\% in this fit.  These events are characterized by the
non-isolation of the lepton and small $\met$, and the fixed rate is
based on comparing to control regions with either low $\met$ or poor
isolation~\cite{l+j}.  The resulting measurement of background rates
in the 0-tag L+J subsample is given in Table~\ref{yields}.  The fit of
the distribution of ANN outputs for this subsample is shown in
Figure~\ref{ann}.
%
%\begin{table}
%\caption{\label{annt} Measurement of signal and background rates in the
%L+J sample from the artificial neural network.  The non-$W$ background
%size is independently estimated.  The uncertainties are combined 
%statistical and systematic.}
%\begin{tabular}{ccccc} \hline \hline
%&
%\multicolumn{1}{c}{\bf All} &
%\multicolumn{1}{c}{\bf 0-tag} &
%\multicolumn{1}{c}{\bf 1-tag} &
%\multicolumn{1}{c}{\bf 2-tag} \\ \hline 
%$t\bar{t}$   & $43.9\pm11.4$ & $16.6\pm8.9$ & $17.2\pm5.1$ & $4.9\pm 1.0$ \\
%$W$+jets     & $52.4\pm11.4$ & $53.4\pm8.9$ & $4.8\pm5.1$ & $0.0^{+1.0}_{-0.0}$ \\
%Non-$W$      & $10.7\pm2.7$ & $9.0\pm2.3$ & $1.0\pm0.3$ & $0.1\pm 0.1$ \\
%Total Background        & $63.1\pm11.7 $ & $62.4\pm9.2$ & $5.8\pm5.1$ & $0.1^{+1.0}_{-0.1}$ \\
%\hline \hline
%\end{tabular}
%\end{table}
%

Systematic uncertainties in the ANN-determined backgrounds are
dominated by our understanding of the jet energy scale, the
renormalization and factorization scale, and the shape of the QCD
template and are strongly anti-correlated between the $t\bar{t}$ and
$W$+jets measurements.  Our ANN-measured $t\bar{t}$ content in the L+J
sample without any tagging requirement is consistent with that found
in our earlier measurement of the $t\bar{t}$ production cross
section~\cite{nnxsec}.  The procedure is repeated in the 1-tag and
2-tag samples, yielding background rates of $5.8 \pm 5.2$ and
$0.1^{+1.0}_{-0.1}$ respectively, consistent with the {\it a priori}
estimates shown in Table~\ref{yields}.  As the {\it a priori}
estimates have smaller uncertainties in the 1-tag and 2-tag
subsamples, the ANN-determined background level is used only for the
L+J 0-tag subsample.

The main backgrounds in the DIL sample are Drell-Yan production
including lepton pairs from the $Z$ resonance, dibosons, and $W$+jets
events with fake leptons.  The total background level of 2.2 $\pm$ 0.6
events in the DIL sample has been estimated elsewhere~\cite{dil}.  The
Drell-Yan rate in $ee$ and $\mu\mu$ events is estimated using
simulated data normalized to the observed rate of $Z$ events in the data.
Other electroweak backgrounds are estimated from MC simulations. The
fake-lepton background is estimated by multiplying each jet in $W$ plus three
or more jet events by a lepton fake rate, measured in complementary jet
samples.
%rates are based on
%measurements in complementary lepton-poor data samples.

Most of the jets in the DIL background events arise from generic QCD
radiation.  To determine the background distribution across the
$i$-tag subsamples, we apply a parameterization of the probability to
tag a generic QCD jet~\cite{l+j}, derived from jet-triggered data
samples, to the jets in the DIL sample, correcting for the enriched
$t\bar{t}$ content of the sample.  The resulting estimates are given
in Table~\ref{yields}; the background in the 2-tag subsample is
negligible.

The $t\bar{t}$ event-tagging efficiency $\epsilon_i$, defined as the
probability to observe $i$ tags in a $t\bar{t}$ event, depends on the
fiducial acceptances for jets that can potentially be tagged, and the
efficiencies to tag those jets~\cite{dima}.  Those efficiencies in
turn depend on the species of the underlying quark in the jet.  The
efficiency $\epsilon_i$ depends strongly on $R$, as $R \neq 1$ implies
fewer $b$ jets available for tagging, and more light-quark jets
available instead.  We use the jet acceptances and tagging
efficiencies to parameterize $\epsilon_i(R)$.  These quantities are
estimated with a sample of simulated $t\bar{t}$ events from the
PYTHIA~\cite{pythia} generator and CDF detector simulation, and their
uncertainties are dominated by our understanding of the control
samples of jet data used to calibrate tagging efficiencies in the
simulation.  The leading determiner of $\epsilon_i$ is the efficiency
to tag a $b$ jet from the decay $t \to Wb$; $\epsilon_b = 0.44 \pm
0.04$ for $b$ jets falling within the fiducial acceptance and having
at least two tracks with silicon information.  The $\epsilon_i$ values
also have small contributions from the efficiencies to tag jets from
$W \to cs$ hadronic decays and from additional QCD radiation in
$t\bar{t}$ events.  The nominal values of $\epsilon_i$ for $R = 1$ are
given in Table~\ref{yields}.  The value of $\epsilon_0$ ($\epsilon_2$)
changes by -0.28 (0.09) as $R$ changes from 0.5 to 1.

The expected event yield in each of the three tagged subsets of each
of the L+J and DIL samples is
\begin{equation}
N_i^\text{exp} = N_\text{inc}^{t\bar{t}} \cdot \epsilon_i(R) + N_i^\text{bkg}\,,
\label{sig}
\end{equation}
where $N_i^\text{bkg}$ is the number of background events in the $i$-tag
subsample and $N_\text{inc}^{t\bar{t}}$ is an estimate of the inclusive
number of $t\bar{t}$ events in the sample, determined by
\begin{equation}
N_\text{inc}^{t\bar{t}} = \sum\nolimits_i (N_i^\text{obs} - N_i^\text{bkg})\,,
\label{sig2}
\end{equation}
where $N_i^\text{obs}$ is the observed number of events in each
subsample.  In this construction, the measured value of $R$ is
independent of any assumption of the overall rate of $t\bar{t}$
production, and is thus sensitive only to the relative
numbers of $t\bar{t}$ events with $i$ tags.

The full likelihood is a product of independent likelihoods for the
L+J and DIL samples.  Each likelihood is a product of Poisson
functions comparing $N_i^\text{obs}$ to $N_i^\text{exp}$ for each value of $i$,
multiplied by Gaussian functions which incorporate systematic
uncertainties in the event-tagging efficiencies and backgrounds,
taking into account the correlations across the different subsamples.
These include correlations in the event-tagging efficiencies through
the single-jet tagging efficiencies; in the common methodology of the
{\it a priori} estimates in the tagged L+J samples; and in the overall
normalization of the DIL backgrounds.  There are a total of five free
parameters in the likelihood to account for these systematic
uncertainties.

\begin{figure}
\includegraphics[width=\columnwidth]{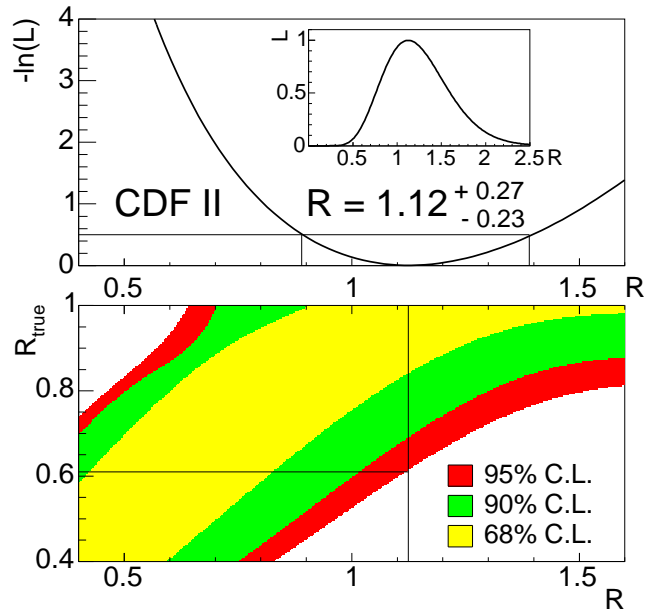}% Here is how to import EPS art
\caption{The upper plot shows the likelihood as a function of $R$ (inset) and
  its negative logarithm. The intersections of the horizontal
  line $\ln(L)=-0.5$ with the likelihood define the statistical $1
  \sigma$ errors on $R$.  The lower plot shows 95\% (outer), 90\%
  (central), and 68\% (inner) CL bands for $R_{true}$ as a function of
  $R$. Our measurement of $R = 1.12$ (vertical line) implies $R >
  0.61$ at the 95\% CL (horizontal line).}
\label{fc}
\end{figure}

The resulting likelihood as a function of $R$ is shown in
Figure~\ref{fc}, along with the negative logarithm of the
likelihood.  We find a central value of
$R~=~1.12^{+0.21}_{-0.19}({\rm stat})^{+0.17}_{-0.13}({\rm syst})$.  The dominant
systematic uncertainties arise from the uncertainty on the background
measurement in the 0-tag L+J sample ($^{+0.14}_{-0.11}$) and from the
overall normalization of the tagging efficiencies
($^{+0.09}_{-0.06}$).  Taken separately, the two final states of
$t\bar{t}$ give consistent results for $R$; the L+J sample alone
yields $R = 1.02^{+0.23+0.21}_{-0.20-0.13}$, and the DIL sample alone
yields $R = 1.41^{+0.46+0.17}_{-0.40-0.13}$.  These $R$ results are
consistent with the SM expectations.
%\begin{figure}[t]
%\includegraphics[scale=.4]{lhf_branching_3}% Here is how to import EPS art
%\caption{Likelihood as a function of $R$ (inset) and its negative logarithm.}
%\label{like}
%\end{figure}

The ratio $R$ can only take on physical values between zero and unity.
We use the Feldman-Cousins prescription~\cite{fc} to set a lower
limit on $R$.  We generate ensembles of pseudo-experiments for
different input values of $R$ ($R_{true}$), and vary the input
quantities of the analysis, {\it e.g.} the background estimates,
taking correlations into account.  Using the likelihood-ratio ordering
principle, we find the acceptance intervals as shown in
Figure~\ref{fc}.  With our measured value of $R$, we find that $R > 0.61$
at the 95\% CL.

Our lower limit on $R$ is the strongest limit on this top-quark
branching ratio to date.  Within the SM, $R =
\frac{|V_{tb}|^2}{|V_{tb}|^2 + |V_{ts}|^2 + |V_{td}|^2}$, up to
phase-space factors.  Assuming three generations and the unitarity of
the CKM matrix, the denominator is unity, and we estimate $|V_{tb}| >
0.78$ at 95\% CL.  All of our measurements of $R$ are consistent with
the SM expectations.

\begin{acknowledgments}
  We thank the Fermilab staff and the technical staffs of the
  participating institutions for their vital contributions. This work
  was supported by the U.S. Department of Energy and National Science
  Foundation; the Italian Istituto Nazionale di Fisica Nucleare; the
  Ministry of Education, Culture, Sports, Science and Technology of
  Japan; the Natural Sciences and Engineering Research Council of
  Canada; the National Science Council of the Republic of China; the
  Swiss National Science Foundation; the A.P. Sloan Foundation; the
  Bundesministerium f\"{u}r Bildung und Forschung, Germany; the Korean
  Science and Engineering Foundation and the Korean Research
  Foundation; the Particle Physics and Astronomy Research Council and
  the Royal Society, UK; the Russian Foundation for Basic Research;
  the Comision Interministerial de Ciencia y Tecnolog\'{i}a, Spain; in
  part by the European Community's Human Potential Programme under
  contract HPRN-CT-2002-00292; and the Academy of Finland.
\end{acknowledgments}

\bibliographystyle{apsrev}
\bibliography{top_R_prl}

\begin{thebibliography}{16}
\bibitem[{geo()}]{geom}
\bibinfo{note}{In the CDF geometry, $\theta$ is the polar angle with respect to
  the proton beam axis, and $\phi$ is the azimuthal angle. The pseudo-rapidity
  is $\eta \equiv -\ln(\tan(\theta/2))$. The transverse momentum, $p_T$, is the
  component of the momentum projected onto the plane perpendicular to the beam
  axis. The transverse energy $E_T$ of a shower or calorimeter tower is
  $E\sin\theta$, where $E$ is the energy deposited.}
\end{thebibliography}

%\bibliography{tlj_secvtx_prd}
%\input{top_R_prl.bbl}

\end{document}